\newcommand{\cm}{{~\rm cm}}
\newcommand{\s}{{~\rm s}}
\newcommand{\km}{{~\rm km}}

\newcommand{\K}{{~\rm K}}
\newcommand{\erg}{{~\rm erg}}
\newcommand{\yr}{{~\rm yr}}

\newcommand{\Gyr}{{~\rm Gyr}}
\newcommand{\pc}{{~\rm pc}}
\newcommand{\kpc}{{~\rm kpc}}
\newcommand{\keV}{{~\rm keV}}
\newcommand{\kev}{{~\rm keV}}

\documentclass[12pt,preprint]{aastex}
\shortauthors{Sternberg \& Soker}

\begin{document}

\title{ENTROPY LIMIT AND THE COLD FEEDBACK MECHANISM IN COOLING FLOW CLUSTERS}

\author{Noam Soker\altaffilmark{1}}

\altaffiltext{1}{Department of Physics,
Technion$-$Israel Institute of Technology, Haifa 32000, Israel;
soker@physics.technion.ac.il.}

\begin{abstract}
I propose an explanation to the finding that star formation and visible filaments strong in
H$\alpha$ emission in cooling flow clusters occur only if the minimum specific entropy
and the radiative cooling time
of the intracluster medium (ICM), are below a specific threshold.
The explanation is based on the cold feedback mechanism.
In this mechanism the mass accreted by the central black hole originates in non-linear
over-dense blobs of gas residing in an extended region of the cooling flow region.
I use the criterion that the feedback cycle period must be longer than the radiative
cooling time of dense blobs for large quantities of gas to cool to low temperature.
The falling time of the dense blobs is parameterized by the ratio of the infall velocity to
the sound speed. Another parameter is the ratio of the blobs' density to that of the
surrounding ICM.
By taking the values of the parameters as in previous papers on the cold feedback model,
I derive an expression that gives the right value of the entropy threshold.
Future studies will have to examine in more detail the role of these parameters, and to show
that the observed sharp change in the behavior of clusters across the
entropy, or radiative cooling time, threshold
can be reproduced by the model.
\end{abstract}

\keywords{galaxies: clusters: general ---
cooling flows ---
galaxies: active ---
intergalactic medium ---
X-rays: galaxies: clusters}

\section{INTRODUCTION}

In the last decade it has become clear that the intra-cluster medium (ICM)
in cooling flow (CF) clusters of galaxies and CF galaxies must be heated,
and the heating process should be stabilized by a feedback mechanism
(see review by McNamara \& Nulsen 2007).
However, in many cases the heating cannot completely offset cooling
(e.g., Wise et al. 2004; McNamara et al. 2004; Clarke et al. 2004; Hicks \& Mushotzky 2005;
Bregman et al. 2006; Salome et al. 2008)
and some gas cools to low temperatures and flows inward (e.g., Peterson \& Fabian 2006).
The mass inflow rate is much below the one that would occur without heating, and the flow is
termed a moderate cooling flow (Soker et al. 2001; Soker \& David 2003; Soker 2004).

One of the likely ingredients of the moderate CF model is a cold feedback
process (Pizzolato \& Soker 2005; Soker 2006; {{{ Pizzolato 2007). }}}
In the cold-feedback model the mass accreted by the central black hole originates
in non-linear over-dense blobs of gas residing in an extended region of $r \sim 5 - 30 \kpc$;
these blobs are originally hot, but then cool faster than their environment and sink toward the center.
The mass accretion rate by the central black hole is determined by the cooling time of the ICM,
the entropy profile, and the presence of inhomogeneities (Soker 2006).
Most important, the ICM entropy profile must be shallow for the blobs to reach the center as cold blobs.
This accretion process is different from the commonly assumed accretion mode
in feedback models where the black hole accretes hot gas (hot feedback mechanism)
from its vicinity, e.g., via a Bondi-type accretion flow (e.g. Churazov et al. 2002; Nulsen 2004;
Omma \& Binney 2004; Chandran 2005; Croton et al. 2006; Balmaverde et al. 2008).

Among other things, the cooling of the ICM to low temperatures is inferred from star
formation (e.g., Rafferty et al. 2008),
and from cool filaments via their H$\alpha$ emission (H$\alpha$ filaments).
In a recent paper Cavagnolo et al (2008) found that almost all clusters with
strong H-alpha emission have a central entropy of
$K \equiv kT n_e^{-2/3} \la 30 \keV \cm^2$; here
$k$ is the Boltzmann constant, $T$ is the ICM temperature, and $n_e$ its electron density.
{{{ This relation might be better presented as a relation between star formation (SF)
and the radiative cooling time: no star formation is seen if the radiative cooling time is
$\tau_{\rm cool} \ga 5 \times 10^ 8 \yr$ or the entropy is $K \ga 30 \keV \cm^2$ (Rafferty et al. 2008). }}}

Voit et al. (2008) suggest that this $K-$H$\alpha$ (or SF$-\tau_{\rm cool}$ relation)
results from a competition
between radiative cooling of cooler regions and heat conduction into these regions.
In high entropy ICM the heat conduction manages to prevent thermal instabilities that potentially
could form H$\alpha$ filaments.
For the cooling function dependance on temperature Voit et al. (2008) assumed $\Lambda \propto T^{1/2}$.
However, for most clusters the temperature in the inner region is $T < 3\keV$, and for a composition
above half a solar the cooling function is basically constant in the range $\sim 1.7-3\keV$.
{{{ This introduces a very small change in the numerical value, and is not significant.
But it does show that the dependance on some power of the entropy is an approximate one.
The same holds for the explanation proposed here (see section 3). }}}
More significant is their assumption that the size of the cooling region is of
the order of the radius of the cluster at the location of the region.
The instabilities at such scale are likely to be very fragmented (fractal),
such that the heat conduction is more efficient than what they assume
(because the temperature gradient is larger across the smaller dimension of a filament,
and the surface area for heat conduction is much larger).
Also, if in some regions the magnetic field lines are closed on themselves on a much
smaller scale than the distance to the cluster center $r$, then the heat
conduction is limited to within magnetic flux tubes (Soker et al. 2004; Soker 2004).
Another point of concern is discussed in section 3.

{{{{ A detail study of the competition between radiative cooling of cool filaments, at a temperature
of $\sim 10^4 \K$, and heat conduction from the ICM is presented by Nipoti \& Binney (2004).
They find that cool filaments survive only if the ICM density is high and the ICM
temperature is low, i.e.,  has a low entropy. However, Nipoti \& Binney (2004)
argue that the filaments do not originate from the hot ICM as Voit et al. (2008) consider,
but rather the cool gas originates from other sources, e.g., AGN activity and external infall.  }}}}

Voit et al. (2008) note that the radiative cooling rate of the ICM depends on entropy, and that the
$K-$H$\alpha$ relation might be alternatively related to the short radiative cooling time of
the low-entropy ICM, {{{ hence better termed SF$-\tau_{\rm cool}$ relation (Rafferty et al. 2008). }}}
In this paper I use the competition between radiative cooling time and the infall time of dense blobs
as developed for the cold feedback mechanism by Pizzolato \& Soker (2005), to explain the
$K-$H$\alpha$ {{{ (SF$-\tau_{\rm cool}$) }}} relation.

\section{THE COLD FEEDBACK EXPLANATION}

In the cold feedback mechanism cold dense blobs fall toward the center of the cluster
and feed the central AGN.
Say a region in the cluster is perturbed, and blobs start to cool.
As the first blobs fall and reach the center, an AGN outburst will heat the perturbed region,
and prevent further cooling.
The condition for large quantities of gas to cool is for the feedback cycle to be longer
than the radiative cooling time
\begin{equation}
t_{\rm cyc} \ga t_{\rm cool}.
\label{eq:cond1}
\end{equation}

The cooling time of a blob with a density of $\delta$ times the ambient density, hence
a temperature $\delta$ times lower, is given by
\begin{equation}
t_{\rm cool} \simeq \frac{nk(T/\delta)}{\delta n_en_p \Lambda},
\label{eq:tcool}
\end{equation}
where $n$, $n_e$, $n_p$ are the total, electron, and proton densities of the ICM, respectively,
and $T$ is the ICM temperature.
In equation (\ref{eq:tcool}) the increase in the cooling function $\Lambda$ and in the density
with decreasing temperature of the cooling blob was accounted for.
For the relevant temperature range here, $1.5 \kev \la kT  \la 4 \kev$, and for solar composition
the cooling function is $\Lambda \simeq 2\times 10^{-23} \erg \cm^2 \s^{-1}$ (Gaetz et al. 1988).

For the feedback cycle time I assume the following.
The dense blob falls with its terminal velocity $v_t$.
Pizzolato \& Soker (2005) studied the fall of dense blobs through the ICM, and found
the terminal velocity to be
\begin{equation}
v_t \simeq  60   
\left(\frac{a}{100 \pc} \right)^{1/2}
\left(\frac{g}{10^{-8} \cm \s^{-2}} \right)^{1/2}
\left(\frac{\delta}{3} \right)^{1/2}
\km \s^{-1} ,
\label{eq:vt}
\end{equation}
where $a$ is the radius of a spherical falling blob, $g$ is the cluster's gravitational
field, and the same scaling as theirs was used here.
{{{ In this paper I scale the blob size and density contrast with the
typical values used by Pizzolato \& Soker (2005). }}}
I incorporate the radius of the blob and $\delta$ into a parameter $\eta$, such that
the average inflow velocity of the blob is $\eta C_s=\eta (5kT/3\mu m_H)^{1/2}$,
where $C_s$ is the sound speed of the ICM.
For a cluster temperature of $2 \kev$, a velocity of $50 \km \s^{-1}$ corresponds to $\eta=0.07$.
The feedback cycle period is then
\begin{equation}
t_{\rm cyc} = \frac{r}{\eta C_s}.
\label{eq:tcyc}
\end{equation}

Inserting equations (\ref{eq:tcool})-(\ref{eq:tcyc}) in condition (\ref{eq:cond1})
gives the condition for the formation of a large quantities of cold gas
\begin{eqnarray}
r \ga D_c \equiv \eta C_s  \frac{n k T}{\delta^2 n_e n_p \Lambda}
\qquad \qquad \qquad \qquad \qquad \qquad \qquad \qquad \qquad
\nonumber \\
= 1.1
\left( \frac{\eta} {0.1} \right)
\left(\frac{\delta}{3} \right)^{-2}
\left(\frac{\Lambda}{2\times 10^{-23} \erg \cm^2 \s^{-1}} \right)^{-1}
\left(\frac{K}{10 \keV \cm^2} \right)^{3/2}
\kpc ,
\label{eq:cond2}
\end{eqnarray}
where $K=kT n_e^{-2/3}$ is the entropy.
The above condition implies that for blobs to cool within the cooling radius, the entropy
must be lower than some threshold.

{{{ The power of the entropy in equation (\ref{eq:cond2}) is an approximate one.
The value of $3/2$ was obtained under the assumption that the infall speed of dense blobs
is proportional to the sound speed, and on that the cooling gas is in the
temperature range where the cooling function does not depend on temperature.
The last assumption is a good approximation for most clusters near the threshold
$K\simeq 30 \keV \cm^2$.
In any case, in the cold feedback mechanism what matters is the radiative cooling time.
This is also seems to be the case from observations (Rafferty et al. 2008). }}}

{{{ Condition (\ref{eq:cond2}) is sensitive to the value of $\delta$ and $\eta$.
Here I simply took the same value used by Pizzolato \& Soker (2005).
In obtaining $\eta$ I assumed a blob size of $\sim 100 \pc$, as the scaling used by them,
and for $\delta$ I took the value of an unstable blob used by them.
The parameter $\eta$ has in it the dependance on the blob radius as $a^{1/2}$, and
on the density contasrt as $\delta^{1/2}$. Thereofre, condition (\ref{eq:cond2})
practically depends on $a^{1/2}$ and $\delta^{3/2}$.
The limit value of $1.1 \kpc$ in condition (\ref{eq:cond2}) is similar to the one
obtained by Voit et al. (2008).
With their typical value of the heat conduction suppression factor $f_c=0.2$
their equation (2) has a limit of $1.8 \kpc$.
However, they use a lower value for the cooling function (because they take the
dependance to be $\Lambda \propto T^{1/2}$).
If I take a value for $\Lambda$ as they take, then the two coefficient are almost equal.
Still, the sensitivity of condition (\ref{eq:cond2}) to the parameters $\eta$ and $\delta$
is a somewhat weak point of the proposed explanation.  }}}

The cold feedback mechanism can account also for the finding that the low entropy clusters have
stronger radio emission (Donahue et al. 2005; Cavagnolo et al. 2008).
The same source of blobs that lead to the formation of H$\alpha$ filaments, will feed the
AGN activity.
Cavagnolo et al. (2008) noticed that below the threshold $K < 30 \keV \cm^2$ there is
no correlation between radio power and the central entropy.
They tentatively speculated that this lack of correlation hints that cold-mode accretion
(Pizzolato \& Soker 2005; Hardcastle et al. 2007) might be the dominant process feeding the
AGN.

\section{DISCUSSION AND SUMMARY}

In this paper I showed that the findings of
{{{ Rafferty et al. (2008) that no star formation (SF) is seen if the radiative cooling
time is $\tau_{\rm cool} \ga 5 \times 10^ 8 \yr$, }}}
and of Cavagnolo et al. (2008) that almost all clusters with
strong H-alpha emission have a central entropy of
$K \equiv kT n_e^{-2/3} \la 30 \keV \cm^2$,
might be explained by comparing the cooling time of dense blobs in the ICM (eq. \ref {eq:tcool})
with the response time (cycle time) of the AGN feedback heating.
This time is taken to be equal to the fall time of the dense blobs to the center
(eq. \ref{eq:tcyc}).
Many dense blobs will cool to low temperatures if the response time is longer
than their cooling time (eq. \ref{eq:cond1}). This leads to equation (\ref{eq:cond2})
which is the main result of this paper.
It shows that for blobs to cool within the cooling flow (CF) radius, the entropy
must be lower than some threshold.

The feedback heating of the ICM in this model is maintained by dense blobs that are accreted
by the central black hole and originate in non-linear over-dense blobs of gas
residing in an extended region of $r \sim 5 - 30 \kpc$;
this is the cold feedback mechanism (Pizzolato \& Soker 2005; Soker 2006).

Criterion (\ref{eq:cond2}) can be compared with the one derived by Voit et al. (2008)
\newline
$r \ga 4 (K/ 10 \keV \cm^2)^{3/2} f_c^{1/2}$, where $f_c$ is the suppression factor of the heat
conductivity, taken to be $f_c \sim 0.1-1$ in their model.
The similarity of the expressions are interesting. They have the same dependance on the entropy,
a similar numerical factor, and some parameters.
Voit et al. (2008) consider their expression to satisfactorily explain their observations.
The similarities between the two conditions show that the cold feedback explanation proposed
here should be considered in future studies as well.
In particular, the result is satisfactory if we consider the crude derivation
performed here, and the inhomogeneous nature of the ICM, where large regions with lower entropy
than the average exist. The dense blobs will be more likely to develop in these regions.

The value of $f_c \sim 0.1-1$ and the large thermally unstable regions considered by
Voit et al. (2008) implies that heat conduction is globally important in the CF region.
However, models based only on global heat conduction are unstable and require fine tuning
(Bregman \& David 1988; Soker 2003; Kim \& Narayan  2003).
The explanation of Voit et al. (2008) requires a stabilizing mechanism.
In a recent paper Guo et al. (2008) built a feedback model based on both heat
conduction and AGN heating. They showed that the AGN heating stabilizes the feedback mechanism.

There is a price payed to achieve the stability in the model of Guo et al. (2008).
(1) The inner boundary of their simulation is at $r=1\kpc$.
They require that all the mass that enters the $r=1\kpc$ sphere will reach the central BH, and the
mechanical energy of the launched jet be $L_{\rm agn}=\epsilon \dot M (1\kpc) c^2$, with
$\epsilon \simeq 0.1-0.3$. This is an extremely efficient conversion of
mass inflowing at $r=1 \kpc$ to mechanical energy of the jets.
(2) In their stable model of A1795 (their model A3) the entropy decreases with radius,
such that the model is unstable to convection.
{{{ This might not be a big concern, as by using a different AGN heating prescription the
negative entropy might be removed (Guo, F., private communication 2008). }}}
(3) In their best models of A2199  (model B3) and of A2052 (model C3) the response time of the
AGN is too long. Using their values of the mass inflow rate and density at $r=1 \kpc$, the
inflow velocity $v_f(1\kpc)$ can be calculated.
The respond of the accreting black hole to changes in the ICM is on a time scale of
$\tau_r={1 \kpc}/v_f$.
 I find $\tau_r= 7 \Gyr$ for model B3 and $\tau_r=8 \Gyr$ for model C3.
 In both cases the response time $\tau_r$ is about an order of magnitude longer than the cooling
 time in the center of these clusters (0.6 and 1.1 Gyr, respectively).
This shows that the AGN heating has no time to respond to changes in the thermal
state of the ICM.
{{{ It is possible that this problem will disappear if an episodic AGN heating
will be applied (Guo, F., private communication 2008). }}}

The explanation proposed here for the $K-$H$\alpha$ relation (or SF$-\tau_{\rm cool}$ relation)
has to overcome some difficulties as well.
In the first step it will be required to show that reasonable values for the two parameters
used in the model ($\delta$ and $\eta$) can lead to the sharp transition from bright
H$\alpha$ clusters, for $K < 30 \keV \cm^2$, to clusters with no, or very weak , detection
of H$\alpha$ emission.
Here I note that the values of $\eta$ and $\delta$ used here are the typical values used
by Pizzolato \& Soker (2005). If I insert the threshold value of $K = 30 \keV \cm^2$ found by
Cavagnolo et al. (2008),
{{{ and $\delta \simeq 2.5-3$ used by Pizzolato \& Soker (2005), }}}
in equation (\ref{eq:cond2}) I find the numerical value there to be
$\sim 13-6 \kpc$. This is inside the range of $r \sim 5-30 \kpc$ considered by
Pizzolato \& Soker (2005) to be the region where dense blobs originate.
In a later step, numerical simulations of dense blobs in the ICM should show that the formation
of cold clouds that are the source of H$\alpha$ emission depends on entropy in the observed way.

{{{ The main difference between the explanation proposed here and that of Voit et al. (2008)
is the importance of the global heat conduction. In recent years the debate on the question of
whether global heat conduction in CF clusters is important or not has intensified.
If that debate is resolved toward a significant global heat conduction, then the explanation
proposed here must be ruled out. If it will turned out that global heat conduction is
not important, then the explanation of Voit et al. (2008) must be ruled out.
In the cold feedback mechanism cold blobs feed the central BH. Therefore, in CF clusters
with strong AGN activity, some cold blobs exist close to the center. Therefore, the prediction here
is that very low level of H$\alpha$ emission might be detected even in CF clusters with
high entropy. }}}

\acknowledgments
{{{  I thank Mark Voit, Brian McNamara, Fabio Pizzolato, and Fulai Guo for helpful comments.
This research was supported in part by the Asher Fund for Space Research at the Technion.
}}}

\end{document}